\documentclass[aps,showpacs,twocolumn,amsmath,amssymb,superscriptaddress,prl]{revtex4-1}

\usepackage{graphicx}
\usepackage{bm}

\usepackage{color}
\usepackage{epstopdf}
\usepackage{amsmath,amssymb}

\begin{document}


\title{Intrinsic Spin Decay Length in Antiferromagnetic Insulator}

\author{Hiroto Sakimura}
\affiliation{School of Materials and Chemical Technology, Tokyo Institute of Technology, Tokyo 152-8552, Japan}
\affiliation{Department of Applied Physics and Physico-Informatics, Keio University, Yokohama 223-8522, Japan}

\author{Akio Asami}
\affiliation{Department of Applied Physics and Physico-Informatics, Keio University, Yokohama 223-8522, Japan}

\author{Takashi Harumoto}
\affiliation{School of Materials and Chemical Technology, Tokyo Institute of Technology, Tokyo 152-8552, Japan}

\author{Yoshio Nakamura}
\affiliation{School of Materials and Chemical Technology, Tokyo Institute of Technology, Tokyo 152-8552, Japan}

\author{Ji Shi}
\affiliation{School of Materials and Chemical Technology, Tokyo Institute of Technology, Tokyo 152-8552, Japan}

\author{Kazuya Ando\footnote{Correspondence and requests for materials should be addressed to ando@appi.keio.ac.jp}}
\affiliation{Department of Applied Physics and Physico-Informatics, Keio University, Yokohama 223-8522, Japan}

\begin{abstract}
We report intrinsic spin decay length of an antiferromagnetic insulator. We found that at an antiferromagnetic/ferromagnetic interface, a spin current generated by spin pumping is strongly suppressed by two-magnon scattering. By eliminating the two-magnon contribution, we discovered that the characteristic length of spin decay in NiO changes by two-orders of magnitude through the paramagnetic to antiferromagnetic transition. The spin decay length in the antiferromagnetic state is longer than 100 nm, which is an order of magnitude longer than previously believed. These results provide a crucial piece of information for the fundamental understanding of the physics of spin transport.  
\end{abstract}

\maketitle

Spintronics relies on the transport of spins in condensed matter~\cite{Zutic,Maekawa,oxfordspin}. Spin transport has been investigated in a variety of materials, including metals, semiconductors, and insulators. In metals and semiconductors, spins are transported by the diffusion of conduction electrons~\cite{oxfordspin}. In contrast, in magnetically-ordered materials, spins can be transported even in the absence of conduction electrons; spins are carried by the elementary excitations of magnetic moments, magnons~\cite{kajiwara}. The magnonic spin current in insulators is of particular recent interest because this sets a new direction for experimental and theoretical studies of the physics of spin transport~\cite{cornelissen2015long,wesenberg2017long}.

Antiferromagnetic insulators is a new class of materials for spin transport~\cite{jungwirth2016antiferromagnetic,JUNGFLEISCH2018865,baltz2018antiferromagnetic}. This class of materials potentially entails a number of advantages as compared to ferromagnets: antiferromagnets are robust against external magnetic fields, produce no stray fields, and display ultrafast dynamics. Since the first observation of the transmission of spins through an antiferromagnetic insulator NiO~\cite{wang2014antiferromagnonic,hahn2014conduction,wang2015spin}, intense experimental and theoretical efforts have been invested in unraveling the physics of the spin transport in antiferromagnetic insulators~\cite{wang2014antiferromagnonic,hahn2014conduction,wang2015spin,doi:10.1063/1.4918990,PhysRevB.92.020409,PhysRevLett.115.266601,lin2016enhancement,takei2014superfluid,PhysRevB.93.054412,bender2017enhanced,qaiumzadeh2017spin,lebrun2018tunable,yuan2018experimental}. In antiferromagnetic insulators, the spin-decay length is known to be typically limited to only a few nanometers~\cite{baltz2018antiferromagnetic}, although theories predict long-distance spin transport in antiferromagnets~\cite{khymyn2016transformation}. This is in stark contrast to the situation for ferromagnetic insulators, where long-distance spin propagation has been observed~\cite{kajiwara,cornelissen2015long}.

In this Letter, we reveal the intrinsic character of magnonic spin transport in an antiferromagnetic insulator. We found that, in the conventional spin-injector/antiferromagnetic-insulator/spin-detector structure, the spin-transmission signal is strongly suppressed by two-magnon scattering. By eliminating the two-magnon contribution in the spin-transmission signal, we show that the spin decay length of a prototypical antiferromagnetic insulator NiO changes by two-orders of magnitude through the paramagnetic to antiferromagnetic transition. This result shows that the intrinsic spin decay length of the antiferromagnetic NiO is an order of magnitude longer than the previously believed, providing an important information for the fundamental understanding of antiferromagnetic spintronics.

\begin{figure}[tb]
	\center\includegraphics[scale=1]{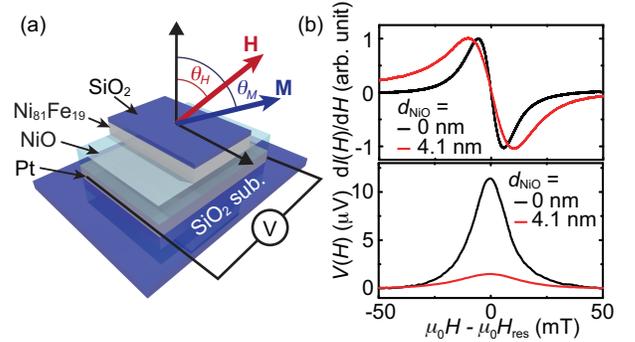}
	\caption{(a) A schematic illustration of the Ni$_{81}$Fe$_{19}$/NiO/Pt trilayer. ${\bf H}$ denotes the external magnetic field. ${\bf M}$ and $\theta_M$ represent the equilibrium direction of the magnetization when ${\bf H}$ is applied at an angle of $\theta_H$ from the film normal. (b) Magnetic field $H$ dependence of the microwave absorption signal $dI/dH$ and voltage signal $V$ for the Ni$_{81}$Fe$_{19}$/NiO/Pt trilayers with $d_{\mathrm{NiO}}=0$~nm (black) and 4.1~nm (red). $I$ is the microwave absorption intensity and $H_\mathrm{res}$ is the FMR field.}
	\label{fig1} 
\end{figure}

\begin{figure*}[tb]
	\center\includegraphics[scale=1]{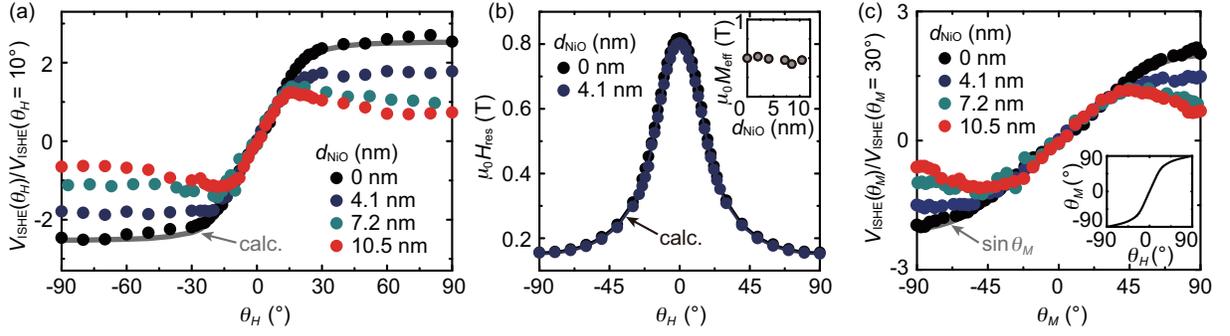}
	\caption{(a) Out-of-plane magnetic field angle $\theta_H$ dependence of the ISHE voltage $V_\mathrm{ISHE}$ for the Ni$_{81}$Fe$_{19}$/NiO/Pt trilayers with $d_\mathrm{NiO}=0$ nm, 4.1 nm, 7.2 nm, and 10.5 nm. The solid curve is the theoretical prediction based on the model of the spin pumping and ISHE, $j_\mathrm{s}(\theta_M)\sin\theta_M$. (b) $\theta_H$ dependence of the FMR field $H_{\mathrm{res}}$ for $d_{\mathrm{NiO}}=$~0 and 4.1~nm. The solid curves are the fitting results. The inset sows the $d_{\mathrm{NiO}}$ dependence of the effective demagnetization field $M_\text{eff}$. (c) Out-of-plane angle of the magnetization-precession axis, $\theta_M$, dependence of  $V_\mathrm{ISHE}$ for the Ni$_{81}$Fe$_{19}$/NiO/Pt trilayers. The solid curve is a function proportional to $\sin\theta_M$. The inset shows $\theta_H$ dependence of $\theta_M$ for $d_\mathrm{NiO}=0$. }
	\label{fig2} 
\end{figure*}

To quantify the intrinsic spin decay length of NiO, we prepared Ni$_{81}$Fe$_{19}$(8)/NiO($d_{\mathrm{NiO}}$)/Pt(5) trilayers on thermally oxidized Si substrates by RF magnetron sputtering at room temperature [see Fig.~\ref{fig1}(a)]. The numbers in brackets represent the thickness of each layer in nm unit, where $d_{\mathrm{NiO}}=0$ to 10.5~nm. The Ni$_{81}$Fe$_{19}$ layer, capped by 4-nm-thick SiO$_2$, is a $1\times 1.5$ mm$^2$ rectangular shape. 
For the Ni$_{81}$Fe$_{19}$/NiO/Pt trilayers, we measured the spin pumping by varying a magnetic field ${\bf H}$ applied at an angle of $\theta_H$ from the film normal at room temperature [see Fig.~\ref{fig1}(a)]. The spin pumping from the Ni$_{81}$Fe$_{19}$ layer injects a spin current into the NiO layer~\cite{tserkovnyak2002spin}. The spin current reaching the Pt layer is converted into an electric voltage $V_\text{ISHE}$ through the inverse spin Hall effect (ISHE) in the Pt layer~\cite{Saitoh}, and thus the spin-current decay in the NiO layer can be characterized by measuring the $d_\mathrm{NiO}$ dependence of  $V_\text{ISHE}$. In Fig.~\ref{fig1}(b), we show the $H$ dependence of the microwave absorption intensity $I(H)$ and voltage $V(H)$ signals for the Ni$_{81}$Fe$_{19}$/NiO/Pt trilayers with $d_{\mathrm{NiO}}=0$ and 4.1~nm at  $\theta_{{H}}=90^\circ$. For the measurement, the Ni$_{81}$Fe$_{19}$/NiO/Pt trilayer was placed at the center of a TE$_{011}$ cavity with the frequency of $f=9.43$~GHz and power of $P=200$ mW, and we measured dc electric voltage $V$ between electrodes attached to the edges of the film [see Fig.~\ref{fig1}(a)]. 
Figure~\ref{fig1}(b) shows that the ISHE voltage $V_\text{ISHE}$ is generated around the FMR field $H=H_\mathrm{res}$. This result also shows that $V_\text{ISHE}\equiv V(H_\mathrm{res})$ is strongly suppressed by inserting the NiO layer, as expected for the spin-current decay in the antiferromagnet. 

Our finding is that magnetic-field angle $\theta_H$ dependence of $V_\text{ISHE}$ strongly depends on the NiO thickness $d_\mathrm{NiO}$. In Fig.~\ref{fig2}(a), we show the $\theta_H$ dependence of $V_\text{ISHE}$ for the Ni$_{81}$Fe$_{19}$/NiO/Pt trilayers with various $d_\mathrm{NiO}$. This result shows that the $\theta_H$ dependence of $V_\text{ISHE}$ for the trilayers with different $d_\mathrm{NiO}$ is the same only around $\theta_H=0$. Here, the variation of $V_\mathrm{ISHE}$ for the film with $d_\mathrm{NiO}=0$ nm is consistent with the standard model of the spin pumping and ISHE~\cite{andoJAPfull}. In this model, when the magnetic damping constant $\alpha$ is independent of $\theta_H$, the spin current generated by the spin pumping is expressed as~\cite{andoJAPfull}
 \begin{equation}
{j_\mathrm{s}}(\theta_M)=\frac{gh^2\hbar\gamma^2\omega }{4\pi\alpha^2A(\theta_M)\left[(4\pi M_\mathrm{s})^2\gamma^2\sin^4\theta_M+4\omega^2\right]},\label{spinc}
\end{equation}
where $g_\text{eff}^{\uparrow\downarrow}$ is the effective spin-mixing conductance, $h$ is the microwave magnetic field, $\gamma$ is the gyromagnetic ratio, $M_\mathrm{s}$ is the saturation magnetization, and $\omega=2\pi f$. $\theta_M$ is the out-of-plane angle of the magnetization-precession axis [see Fig.~\ref{fig1}(a)]. $A(\theta_M)={2\omega}\left[{4\pi M_\mathrm{s}\gamma \sin^2\theta_M+\sqrt{(4\pi M_\mathrm{s}\gamma)^2 \sin^4\theta_M+4\omega^2}}\right]^{-1}$ is the precession ellipticity factor. When the magnetization-precession axis is oblique to the film plane, the ISHE voltage $V_\mathrm{ISHE}$ is proportional to ${j_\mathrm{s}}(\theta_M)\sin\theta_M $ because of ${\bf j}_\mathrm{s}^\mathrm{Pt}\parallel {\bf j}_\mathrm{c}^\mathrm{Pt}\times {\bm \sigma}$~\cite{andoJAPfull}, where $j_\mathrm{s}^\mathrm{Pt}$ is the spin current density injected into the Pt layer and $j_\mathrm{c}^\mathrm{Pt}$ is the charge current density generated by the ISHE. $\bm{\sigma}$ is the spin-polarization direction of the spin current, which is parallel to the magnetization-precession axis. As shown in Fig.~\ref{fig2}(a), this model well reproduces the experimental data only for $d_\mathrm{NiO}=0$ nm [see the solid curve]. For the calculation, we determined $\theta_M$ and $M_\mathrm{s}$ from measured $\theta_H$ dependence of $H_\mathrm{res}$, shown in Figs.~\ref{fig2}(b), by solving $\omega={\gamma}\sqrt{H_XH_Y}$, where $\omega= 2\pi f$, $H_X=H_\mathrm{res}\cos(\theta_H-\theta_M)-M_{\mathrm{eff}}\cos^2\theta_M$, and $H_Y=H_\mathrm{res}\cos(\theta_H-\theta_M)-M_{\mathrm{eff}}\cos(2\theta_M)$~\cite{mizukami2002effect,lindner2009two,landeros2008two,arias1999extrinsic,arias2000extrinsic} [see the inset to Figs.~\ref{fig2}(b) and \ref{fig2}(c)]. $M_\mathrm{eff}\simeq M_\mathrm{s}$ is the effective demagnetization field.

To clarify the origin of the anomaly in the $\theta_H$ dependence of $V_\mathrm{ISHE}$ for the Ni$_{81}$Fe$_{19}$/NiO/Pt trilayers with $d_\mathrm{NiO}\neq 0$ nm, we plot $\theta_M$ dependence of $V_\mathrm{ISHE}$ in Fig.~\ref{fig2}(c). Since $j_\mathrm{c}^\mathrm{Pt}(\theta_M)$ does not change drastically with $\theta_M$, $V_\text{ISHE}$ is approximately proportional to $\sin \theta_M$. In fact, the $\theta_M$ dependence of $V_\text{ISHE}$ is consistent with this scenario for the Ni$_{81}$Fe$_{19}$/Pt bilayer ($d_\mathrm{NiO}=0$ nm). However, for the Ni$_{81}$Fe$_{19}$/NiO/Pt trilayers, the measured $V_\mathrm{ISHE}$ values are proportional to $\sin\theta_M$ only at $|\theta_M|<45^\circ$ as shown in Fig.~\ref{fig2}(c); $V_\mathrm{ISHE}$ deviates from $\sin\theta_M$ at $|\theta_M|>45^\circ$ with increasing the thickness of the NiO layer.

\begin{figure}[tb]
	\center\includegraphics[scale=1]{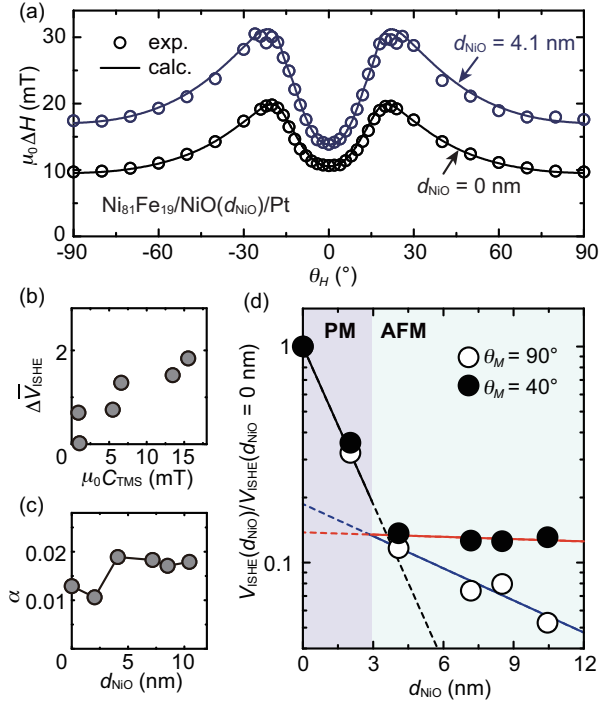}
	\caption{(a) $\theta_H$ dependence of the peak-to-peak FMR linewidth $\Delta H$ for $d_{\text{NiO}}=0$ nm and 4.1 nm. The open circles are the experimental data and the solid curves are the fitting results~\cite{supplementary}. (b) The relation between the amplitude of the two-magnon scattering $C_\mathrm{TMS}$ and $\Delta \bar{V}_\mathrm{ISHE} \equiv V_\mathrm{ISHE}^\mathrm{calc}(\theta_H=90^\circ)/V_\mathrm{ISHE}^\mathrm{calc}(\theta_H=10^\circ)-V_\mathrm{ISHE}^\mathrm{exp}(\theta_H=90^\circ)/V_\mathrm{ISHE}^\mathrm{exp}(\theta_H=10^\circ)$, where $V_\mathrm{ISHE}^\mathrm{calc(exp)}(\theta_H)$ is the calculated(measured) ISHE voltage at $\theta_H$. (c) $d_{\text{NiO}}$ dependence of the magnetic damping constant $\alpha$. (d) $d_{\text{NiO}}$ dependence of $V_{\text{ISHE}}$ at ${\theta_M = 90^\circ}$ (open circles) and $\theta_M = 40^\circ$ (solid circles). The solid line in black is the fitting result using an exponential function, $\exp(-d_\mathrm{NiO}/\lambda_\mathrm{NiO})$, for $d_\mathrm{NiO}<3$ nm. The solid lines in blue and red are the fitting result for the data at $\theta_M=90^\circ$ and $\theta_M=40^\circ$ for $d_\mathrm{NiO}>3$ nm, respectively. }
	\label{fig3} 
\end{figure}

The drastic change in $V_\mathrm{ISHE}$ at $|\theta_M|>45^\circ$ indicates that the nontrivial variation of $V_\mathrm{ISHE}$ is caused by two-magnon scattering in the Ni$_{81}$Fe$_{19}$/NiO/Pt trilayers. The two-magnon scattering can be induced only when $|\theta_M|>45^\circ$ because the degenerated states with $k=0$ mode disappear at $|\theta_M|<45^\circ$~\cite{kurebayashi2013uniaxial,arias1999extrinsic,arias2000extrinsic}. Here, as shown in Fig.~\ref{fig1}(b), the peak-to-peak FMR linewidth $\Delta H$ is clearly enhanced by inserting the NiO layer, despite the negligible change in the effective demagnetization field $M_\mathrm{eff}$ [see the inset to Fig.~\ref{fig2}(b)]. To quantitatively study the damping enhancement induced by the NiO insertion, we plot $\theta_H$ dependence of $\Delta H$ in Fig.~\ref{fig3}(a). Figure~\ref{fig3}(a) shows $\Delta H(\theta_H=\theta_M=0^\circ)\simeq \Delta H(\theta_H=\theta_M=90^\circ)$ for $d_\mathrm{NiO}=0$ nm, while $\Delta H(\theta_H=\theta_M=0^\circ)<\Delta H(\theta_H=\theta_M=90^\circ)$ for $d_\mathrm{NiO}=4.1$ nm. This result indicates that $\Delta H(\theta_H=90^\circ)$ for the Ni$_{81}$Fe$_{19}$/NiO/Pt trilayer is influenced by the two-magnon scattering.

The two-magnon scattering is known to be activated by the random fluctuation of uniaxial anisotropy, surface/interface roughness, and defects~\cite{azevedo2000extrinsic,arias1999extrinsic,arias2000extrinsic,landeros2008two,lindner2009two}. We note that in the Ni$_{81}$Fe$_{19}$/NiO/Pt trilayers, the NiO layer is polycrystalline, as evidenced by the X-ray diffractometry~\cite{supplementary}. This suggests that the two-magnon scattering can be induced by the random fluctuation of uniaxial anisotropy due to randomly oriented exchange bias fields~\cite{sakimura2018}. 
In fact, the measured $\theta_H$ dependence of $\Delta H$ is well reproduced by a calculation which takes into account the additional damping due to the two-magnon scattering as shown in Fig.~\ref{fig3}(a)~\cite{sakimura2018,lindner2009two} [for details, see \cite{supplementary}]. In the Ni$_{81}$Fe$_{19}$/NiO/Pt trilayers, the random fluctuation of uniaxial anisotropy due to the randomly oriented exchange bias increases with $d_\mathrm{NiO}$~\cite{sakimura2018}; although the surface roughness of the NiO layer is almost unchanged with $d_\mathrm{NiO}$~\cite{supplementary}, the amplitude of the two-magnon scattering $C_\mathrm{TMS}$ increases with $d_\mathrm{NiO}$, which is reminiscent of the increased suppression of $V_\mathrm{ISHE}$ with $d_\mathrm{NiO}$ shown in Fig.~\ref{fig2}(c). 
Here, we characterize the suppression of $V_\mathrm{ISHE}$ induced by the NiO insertion as the difference between the measured $V_\mathrm{ISHE}$ and $V_\mathrm{ISHE}$ calculated using the conventional spin-pumping model, $\Delta \bar{V}_\mathrm{ISHE} \equiv V_\mathrm{ISHE}^\mathrm{calc}(\theta_H=90^\circ)/V_\mathrm{ISHE}^\mathrm{calc}(\theta_H=10^\circ)-V_\mathrm{ISHE}^\mathrm{exp}(\theta_H=90^\circ)/V_\mathrm{ISHE}^\mathrm{exp}(\theta_H=10^\circ)$, where $V_\mathrm{ISHE}^\mathrm{calc}(\theta_H)$ and $V_\mathrm{ISHE}^\mathrm{exp}(\theta_H)$ are the calculated and measured ISHE voltage at $\theta_H$, respectively [see Fig.~\ref{fig2}(a)]. 
To clarify the relation between $C_\mathrm{TMS}$ and the voltage suppression, we plot $\Delta \bar{V}_\mathrm{ISHE} $ with respect to $C_\mathrm{TMS}$, extracted by the calculation shown in Fig.~\ref{fig3}(a). As shown in Fig.~\ref{fig3}(b), $\Delta \bar{V}_\mathrm{ISHE} $ increases with $C_\mathrm{TMS}$, supporting that the suppressed $V_\mathrm{ISHE}$ signals at $|\theta_M|>45^\circ$ is caused by the two-magnon scattering.

From the calculation of the $\theta_H$ dependence of $\Delta H$, we also extracted the damping constant $\alpha= \mu _ { 0 } [\Delta H-(\Delta H_\mathrm{inhomo}+\Delta H_\mathrm{TMS})] ( { \sqrt { 3 }  }/ { 2})  ({  \gamma \Xi} /{ \omega })$, where $\Delta H_\mathrm{inhomo}$ and $\Delta H_\mathrm{TMS}$ are the linewidth due to inhomogeneity and two-magnon scattering, respectively. $\Xi$ is the dragging function~\cite{supplementary}. Figure~\ref{fig3}(c) shows that $\alpha$ decreases at $d_{\text{NiO}}=2.0$~nm, while $\alpha$ increases above $d_{\text{NiO}}=4.1$~nm, consistent with previous reports~\cite{wang2014antiferromagnonic, wang2015spin}; $\alpha$ decreases due to the decoupling of the Ni$_{81}$Fe$_{19}$ and Pt layers by the insulating and non-N\'eel-ordered NiO layer because the N\'eel temperature of 2-nm-thick NiO is below the room temperature~\cite{gruyters2002structural,baruth2008enhanced,wang2014spin}. Above $d_{\text{NiO}}=4.1$~nm, $\alpha$ increases because of the enhanced antiferromagnetic correlation due to the thickness growth~\cite{wang2014antiferromagnonic,lang2007dependence}.

Commonly, the spin decay length $\lambda_\mathrm{NiO}$ of NiO is obtained from the thickness $d_\mathrm{NiO}$ dependence of $V_\mathrm{ISHE}$ at $\theta_H=\theta_M=90^\circ$~\cite{wang2014antiferromagnonic, wang2015spin,hahn2014conduction}. Following this procedure, we plot the $d_\mathrm{NiO}$ dependence of $V_\mathrm{ISHE}$ at $\theta_M=90^\circ$ in Fig.~\ref{fig3}(d). This result shows that the spin decay length is increased from  $\lambda_\mathrm{NiO}=1.8$ nm for $d_\mathrm{NiO}<3$ nm to $\lambda_\mathrm{NiO}=8.8$ nm for $d_\mathrm{NiO}>3$ nm. The increase of $\lambda_\mathrm{NiO}$ can be attributed to the paramagnetic to antiferromagnetic transition; for $d_\mathrm{NiO}<3$~nm, the N\'eel temperature is lower than the room temperature, while the NiO layer with $d_{\text{NiO}}>3$~nm is antiferromagnetic at room temperature ~\cite{wang2014antiferromagnonic,baruth2008enhanced,gruyters2002structural}. 
$\lambda_\mathrm{NiO}=8.8$ nm in the antiferromagnetic state is consistent with previous reports~\cite{wang2014antiferromagnonic,wang2015spin}. However, we note that, as is clear from Fig.~\ref{fig2}(a), the $V_\mathrm{ISHE}$ signals at $\theta_H=90^\circ$ are strongly suppressed by the two-magnon scattering. This results in under estimation of the spin decay length in the antiferromagnetic state because the voltage suppression increases with $d_\mathrm{NiO}$.

The intrinsic spin decay length, where the two-magnon contribution is excluded, can be determined only from the $d_\mathrm{NiO}$ dependence of $V_\mathrm{ISHE}$ at $|\theta_M|<45^\circ$, where the voltage suppression due to the two-magnon scattering is absent. As shown in Fig.~\ref{fig3}(d), the $d_\mathrm{NiO}$ dependence of $V_\mathrm{ISHE}$ at $\theta_M=40^\circ$ is clearly different from that at 
$\theta_M=90^\circ$. From the data at $\theta_M=40^\circ$, for the antiferromagnetic NiO, we obtain $ \lambda_\mathrm{NiO}=109$ nm, which is almost ten times longer than previously reported values~\cite{wang2014antiferromagnonic,wang2015spin}. We also note that the characteristic length of spin decay in NiO changes by two-orders of magnitude through the paramagnetic to antiferromagnetic transition, illustrating the crucial role of the antiferromagnetic order for efficient spin transport in antiferromagnetic insulators.


In summary, we investigated magnonic spin transport in an antiferromagnetic insulator NiO. We found that in the in-plane magnetic field geometry, the spin transport signal is strongly suppressed by the two-magnon scattering. By changing the magnetic-field angle, the two-magnon scattering contribution can be eliminated, which enables to determine the intrinsic spin decay length of the antiferromagnetic insulator. 
Although the spin transport signal for the Ni$_{81}$Fe$_{19}$/NiO/Pt trilayer with much thicker $d_\mathrm{NiO}$ is difficult to measure because the surface roughness of the NiO layer increases with $d_\mathrm{NiO}$, our result shows that the intrinsic spin decay length of the prototypical antiferromagnetic insulator NiO is longer than 100 nm, which is an order of magnitude longer than previously believed. The result shows that the spin decay length changes by two-orders of magnitude through the paramagnetic to antiferromagnetic transition. Our results therefore demonstrate the crucial role of the antiferromagnetic order for efficient spin transport in antiferromagnetic insulators, as well as the two-magnon scattering in quantifying the spin transport in antiferromagnets.

\begin{acknowledgments}
This work was supported by JSPS KAKENHI Grant Numbers 26220604, 26103004, the Asahi Glass Foundation, and JGC-S Scholarship Foundation. H.S. is supported by JSPS Grant-inAid for Research Fellowship for Young Scientists (DC1) No. JP17J03624.
\end{acknowledgments}

%

\end{document}